\documentclass[12pt,twoside]{article}
\usepackage{fleqn,espcrc1}

\usepackage{graphicx}
\usepackage[figuresright]{rotating}

\newcommand{\AmS}{{\protect\the\textfont2
A\kern-.1667em\lower.5ex\hbox{M}\kern-.125emS}}

\title{Neutron Electric Form Factor from ${}^3\overrightarrow{\mbox{He}}(\vec{e},e'n)$}

\author{D.J. Boersma\address{NIKHEF, P.O. Box 41882,
                             1009 DB Amsterdam, The Netherlands},
        on behalf of the 9405 collaboration}

\begin{document}

\maketitle

\begin{abstract}

In a double polarised electron scattering experiment data
have been taken from which $G_E^n$ can be extracted for $Q^2=0.2\mbox{ GeV}^2/c^2$.
We have measured asymmetries corresponding to the spin-spin correlation functions $A'_x$
and $A'_z$ and to the target analyzing power $A_y^0$. In order to investigate the
influence of the reaction mechanisms and the different components of the $^3\mbox{He}$
wave function, also the $(e,e'p)$ and $(e,e'd)$ channels have been measured simultaneously.
Some preliminary results are presented.

\end{abstract}

\section{Introduction}

The electric form factor of the neutron has been relatively poorly known for a long time,
which is mainly due to the absence of free neutron targets, so one has to resort to light
nuclei as effective neutron targets. This complicates the interpretation of the experiment,
since nuclear structure and proton contributions need to be taken into account.
E.g. the Platchkov inclusive scattering data on Deuterium \cite{Pla90} have an impressive
statistical accuracy, but the $G_E^n$ extraction depends strongly on the choice
of the $NN$-potential used for the deuteron wave function: the absolute magnitude
varies over 50\%.

Double polarized experiments have been proposed because this model dependency and experimental
systematic uncertainties cancel out to a great extent by measuring asymmetries. Moreover,
in exclusive experiments the results depend less on the proton contributions; but on
the other hand become more sensitive to final state interactions.

The general expression for the cross section of reactions with polarised beam and
target is \cite{Laget}:

\begin{displaymath}
\sigma = {\sigma}_0 \times \left\{ 1 + \mathbf{A^0}\cdot\mathbf{S}
                                     + h \left( A_e + \mathbf{A'}\cdot\mathbf{S} \right) \right\}
\end{displaymath}
where $\mathbf{S}$ and h are the target polarisation and the electron helicity; $\mathbf{A^0}$
and $A_e$ are the target and electron analyzing powers and $\mathbf{A'}$ the spin-spin 
correlation functions.
The Cartesian coordinate frame for the vectorial polarisation observables
has \^{x} in the scattering plane, perpendicular to the 3-momentum transfer $\mathbf{q}$,
\^{y} perpendicular to the scattering plane and \^{z} parallel to $\mathbf{q}$.

On a free neutron (and in a nucleus in PWIA) $A'_x$ for $(\vec{e},e'n)$ is proportional to
$G_E^n/G_M^n$ and $A'_z$ in the inclusive channel is proportional to $G_M^n$, while
all other components of $\mathbf{A^0}$ and $\mathbf{A'}$ are zero.

In our experiment we chose ${}^3\mathrm{He}$ as an effective neutron target.
Since for 90\% of the ${}^3\mathrm{He}$ ground state wave
function the two protons are in a relative
$S$ state, ${}^3\overrightarrow{\mathrm{He}}$ may be considered as an
effective polarised neutron target. However, the $D$ state and the $S'$ state
still contribute about 8\% and 2\%, respectively, which can not be neglected
since their contribution to $A'_x$ is as important as that of $G_E^n$.

Another complicating factor is the influence of the final state interactions.
Their importance was demonstrated by the measurement of $A_y^0$ in the first run of the 9405
experiment in 1997, where this observable, which is zero in PWIA, was found
to be $0.50\pm0.05$ at $Q^2=0.2 \mbox{ GeV}^2/c^2$ \cite{HRP99}. This suggests that
an accurate theoretical description, that takes all relevant reaction reaction
mechanisms consistently into account, is needed for a reliable interpretation
of the data, in particular for the extraction of $G_E^n$.
We compare our data for $(e,e'p)$, $(e,e'd)$ and $(e,e'n)$ with calculations
Laget \cite{Laget,Lag94}, Nagorny \cite{Nagor} and
Golak (Gl\"{o}ckle \emph{et al.} \cite{Glock}).

\section{Experimental setup}

\subsection{Polarised electrons}

The experiment was performed in the internal target hall of the MEA/AmPS
facility at NIKHEF.
Pulses of $2.1\mu\mbox{s}$ of polarised electrons were produced by means of photoemission
from strained layer III-V semiconductor crystals in an ultrahigh vacuum electron gun
\cite{Put98}. The spin of these 100 keV electrons could be directed (using electrostatic
bends and solenoids) and measured (with a Mott polarimeter). The electrons
were accelerated by MEA to 720 MeV and injected into the AmPS storage
ring. A 'Siberian Snake', consisting of a set of superconducting
solenoids, preserved the electron polarisation during storage. The
electron polarisation in the target area was measured with a Compton
Backscattering Polarimeter \cite{Pas99}. It was possible to store over 200 mA, but
we observed that the beam polarisation was lost when more than 140 mA was
injected. During the experiment a maximum injection current of 120 mA was used.

\subsection{Polarised ${}^3\mathbf{He}$}

In the target area, the beam is led through a \O{}~15~mm, 40~cm long cylindrical
storage cell. Polarised ${}^3\mbox{He}$ gas is bled into the storage cell with a flow
of $10^{17} \mathrm{atoms}/s$ resulting in a density of $10^{15} \mathrm{atoms}/\mathrm{cm}^2$
at a temperature of 30 K. The ${}^3\mbox{He}$ gas was polarised in
a pumping cell directly above the target with
a metastability-exchange optical pumping technique. Transitions to the
$2^3S_1$ state were induced with an RF generator, after which illumination with
circularly polarised laser light of 1083 nm excited the atoms to a ${}^2P_{0,1,2}$
state, thereby depleting the $2^3S_1$ hyperfine levels with the nuclear spin (anti-)parallel
to the target holding field.  The polarisation
is then transferred to the atoms in the ground state via metastability-exchange collisions.
The nuclear polarisation was measured by monitoring the fluorescence light of the
$3^1D_2 - 2^1P_1$ transition.

\subsection{Detectors}

The scattered electrons were detected with the BigBite spectrometer
\cite{Lange98}, consisting of a 1~Tesla dipole magnet, two sets
of drift chambers for tracking and a scintillator and \v{C}erenkov
detector for triggering and particle identification. The momentum
bite is 250-900~MeV/c with a resolution of about
$0.5~\%$, while a solid angle extended over $10^\circ$ in the polar angle
(resolution $0.2^\circ$, central angle $40^\circ$) and $34^\circ$
in the azimuthal angle (resolution $0.2^\circ$).

Protons and deuterons were detected with the Range Telescope, consisting
of two wire chambers, a hodoscope and fifteen 1~cm thick scintillator
layers. Protons with an energy of 35-150~MeV and deuterons with an energy
of 45-200~MeV could be detected with a resolution of 1~MeV in energy
and $1^\circ$ in angle.

Neutrons were detected in two consecutive time-of-flight walls,
positioned at 2.2~m and 3.1~m (respectively) from the centre of
the target at a central angle of $56^\circ$ with the electron beam.
Each wall consists of four horizontally stacked telescopes, each
telescope consisting of a $20\times{}20{}\times{}160 \mathrm{cm}^3$
scintillator bar preceded by 1.0~cm and 0.3~cm thick veto layers.
All layers have double-sided readout, enabling position determination
with a resolution of about 5~cm. The time of flight of the neutrons
from the interaction vertex to the neutron detector (ND) was deduced
from the trigger time difference of BigBite and the ND, corrected
for the electron time-of-flight and for the time of flight of the
scintillator light from the impact position to the triggering
photomultiplier. Detection of protons in elastic
$\mathrm{H(}e,e'p\mathrm{)}$ determined the offset in the time-of-flight.
The resolution of 1.0~ns (sigma) was limited by the jitter in the 
triggering scintillator of BigBite.

\begin{figure}[htb]
  \begin{minipage}[t]{75mm}
    \includegraphics[width=70mm]{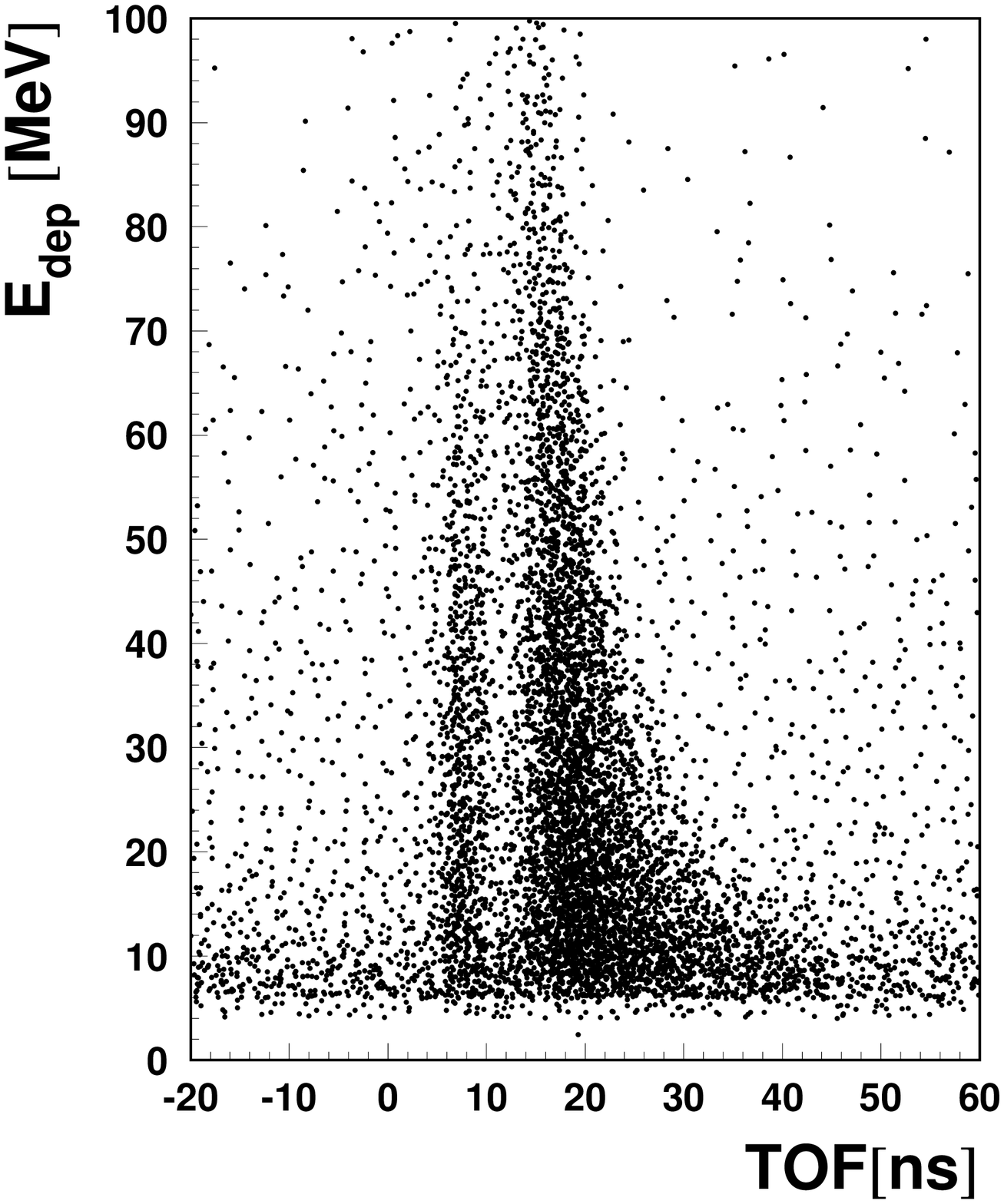}
    \caption{Energy deposit of neutral particles in one of the ND bars versus time of flight.
             For this plot, $\omega$ ranges from 20 till 470 MeV. For $\omega<170$ the pion
             band at 8~ns disappears. After putting a cut on $E_m$ most of the random
             background vanishes.}
    \label{Azeps}
  \end{minipage}
  \hspace{\fill}
  \begin{minipage}[t]{75mm}
    \includegraphics[width=70mm]{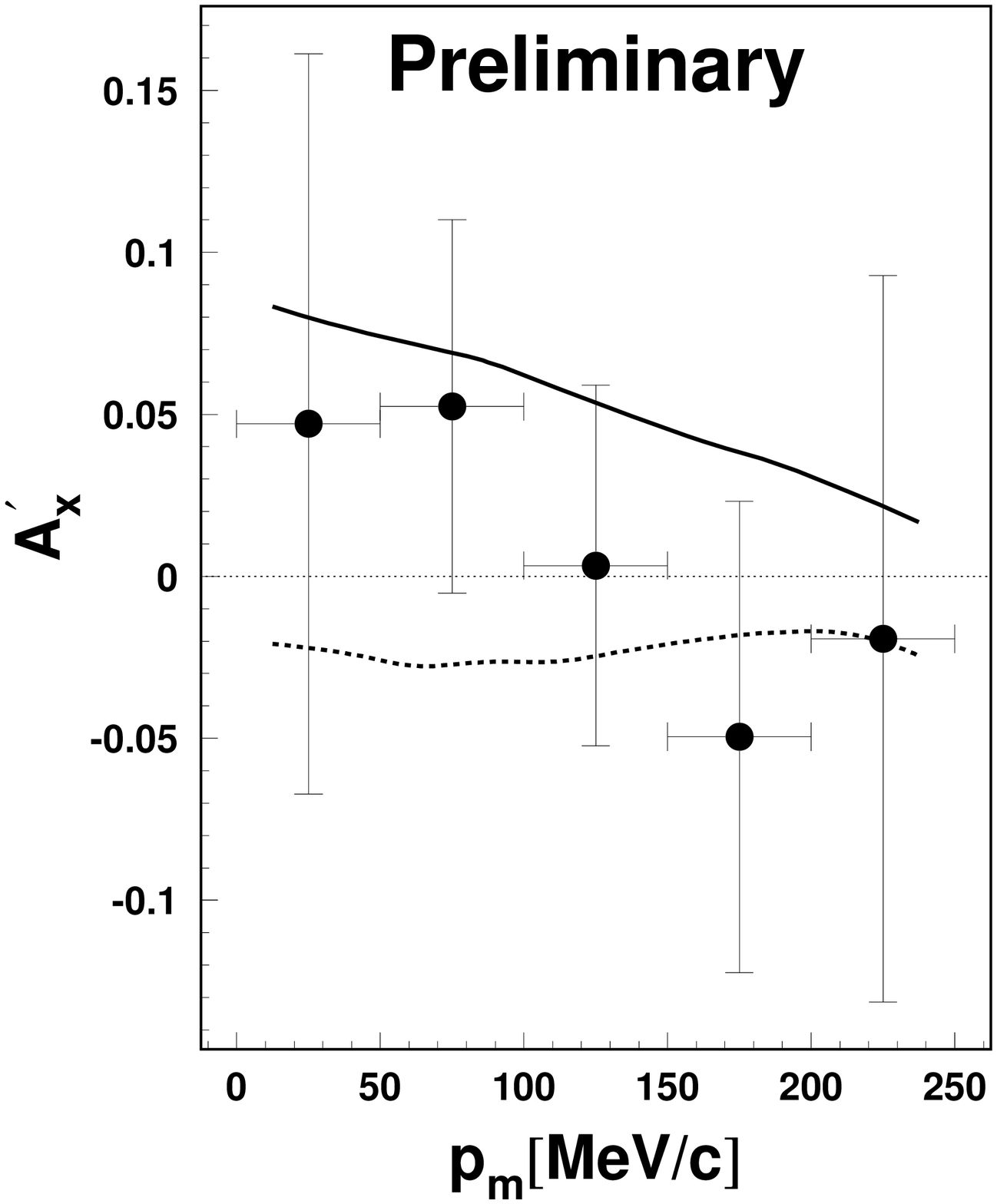}
    \caption{$A'_x$ as function of the missing momentum. The curves represent
             one-loop calculations of Nagorny, averaged over the total acceptance
             with a Monte Carlo program. The solid curve was calculated with $G_E^n$ equal
             to the Galster fit \cite{Galster}, the dotted curve with $G_E^n=0$. Both the
             data and the curves are preliminary.}
    \label{Axeps}
 \end{minipage}
\end{figure}

Neutral particles were selected by requiring no hit in the veto layers
of the triggering scintillator bar and those of the neighbouring bars. A
plot of deposited energy versus time for neutral particles is shown in
figure~\ref{Azeps}. The band at 8~ns is associated with pion production.
It disappears when $\omega<170\mbox{ MeV}$ is required. Since the
random coincidence events have an almost flat distribution in $E_m$
over a range of severel hundred MeV, most of this background can be rejected
by placing a cut on $E_m$.

\section{Results}

In figure~\ref{Axeps} $A'_x$ is plotted versus the missing momentum
($p_m=|\mathbf{q}-\mathbf{p}_n|$), all other kinematical degrees of freedom
are averaged over the acceptance, within the following ranges: 
\begin{displaymath}
  \begin{array}{ccccc}
0.14 \mbox{ GeV}^2 & < & Q^2    & < & 0.26 \mbox{ GeV}^2 \\
  40 \mbox{ MeV}   & < & \omega & < &  170 \mbox{ MeV} \\
 550 \mbox{ MeV}   & < & E'     & < &  680 \mbox{ MeV} \\
 -10 \mbox{ MeV}   & < & E_m    & < &   40 \mbox{ MeV} \\
 200 \mbox{ MeV}   & < & p_n    & < &  600 \mbox{ MeV} \\
  \end{array}
\end{displaymath}
with $E_m=\omega-T_n-T_{pp}$, where $T_n$ is the neutron kinetic energy and
$T_{pp}=\sqrt{p_m^2+4m_p^2}-2m_p$ is the kinetic energy of the center of mass
of the recoiling two protons. The resolution in $E_m$ was 13 MeV.

Figure~\ref{Axeps} contains about 50-70\% of the final statistics. The analysis
is ongoing and this result is very preliminary. One-loop calculations of Nagorny were
performed for a grid of 50000 points, which was used by a Monte Carlo program for
interpolation. The resulting curves for $G_E^n=0$ and $G_E^n$ equal to the Galster fit
are plotted seem to suggest that this data set will result in a value between zero and
one times the Galster fit. The two-loop calculations differ from the one-loop
calculations at low $Q^2$; this needs to be investigated further.

\pagebreak

\end{document}